\documentclass[a4paper]{article}
\usepackage{ISCSLP2026}
\usepackage{ifthen}
\usepackage{xcolor}
\usepackage{array}
\usepackage{booktabs}
\usepackage{url}

\usepackage{enumitem}

\usepackage{hyperref}

\usepackage{graphicx}
\usepackage{multirow}

\usepackage{booktabs}
\usepackage{tabularx}
\usepackage{array}

\newcommand{\nvvtag}[1]{\texttt{[#1]}}
\title{Source-Adaptive Data Curation for Bilingual NVV-Aware ASR}

\name{
  Yuang Cao$^1$, Qirui Zhan$^1$, Jingbin Hu$^1$, Ziyu Zhang$^1$,
  Yunxiang Chen$^2$, Houdun Liu$^2$, Su Feng$^2$, Bengu Wu$^3$,
  Lei Xie$^{1,\star}$, Liumeng Xue$^{4,5,\star}$\thanks{$^{\star}$Corresponding author.}
}
\address{
$^1$~ASLP@NPU, Northwestern Polytechnical University, China\\
$^2$~Shenzhen Pimei Technology, China\\
$^3$~Yutuzhineng, China\\
$^4$~State Key Laboratory of Novel Software Technology, Nanjing University, Nanjing, China\\
$^5$~School of Intelligence Science and Technology, Nanjing University, Suzhou, China
}

\email{
  yacao@mail.nwpu.edu.cn, lxie@nwpu.edu.cn, lmxue@nju.edu.cn
}

\begin{document}
\maketitle

\begin{abstract}
\vspace*{-5pt}
Nonverbal vocalizations (NVVs), such as laughter, sighs, breaths, and coughs, convey affective and interactional information that conventional automatic speech recognition (ASR) systems often discard. We present a bilingual Mandarin--English system for Track~1 of the NVVSpeech Challenge at ISCSLP~2026, which requires joint transcription of lexical content and 16 NVV categories at their transcript-relative positions. Our NVV-Aware Whisper adapts Whisper-medium through checkpoint-compatible vocabulary remapping, enabling lexical tokens and inline NVV tags to be decoded within a unified autoregressive sequence without expanding the vocabulary. To provide reliable and diverse supervision, we further introduce a source-adaptive data curation strategy that refines public NVV corpora through acoustic augmentation and multimodal LLM filtering, while mining spontaneous NVVs from in-the-wild media through automated preprocessing and annotation. Under the official bilingual evaluation protocol, the proposed system improves final score from 33.32 to 53.61, with ablations confirming the complementary benefits of the proposed data-curation components.

\end{abstract}

\section{Introduction}

Beyond lexical content, spoken interaction contains rich nonverbal vocalizations (NVVs), such as laughter, sighs, breaths, coughs, and gasps, which convey affective, interactional, and physiological information~\cite{DBLP:conf/icassp/GongYG22,ye2025scalable,liao2025nvspeech,mai2026mnv,schuller2013paralinguistics}. Conventional ASR systems typically omit or collapse these events into generic non-speech symbols, losing information important for dialogue understanding, media analysis, and computational paralinguistics~\cite{yang2026beyond}. NVV-aware ASR instead represents lexical content and inline NVV events in one transcript~\cite{inaguma2018end,shione2023automatic}, requiring correct category and transcript-relative placement; this is harder in bilingual Mandarin--English settings, where speaking styles, acoustic realizations, and annotation distributions differ across languages and sources.

Recent datasets have expanded coverage of nonverbal speech, including ESC~\cite{DBLP:conf/mm/Piczak15}, VocalSound~\cite{DBLP:conf/icassp/GongYG22}, NonVerbalSpeech-38K~\cite{ye2025scalable}, NVSpeech~\cite{liao2025nvspeech}, MNV-17~\cite{mai2026mnv}, SMIIP-NVV~\cite{DBLP:conf/mm/WuL0WLJBZL25}, SynParaSpeech~\cite{bai2026synparaspeech}, and NonverbalTTS~\cite{borisov2025nonverbaltts}. However, existing resources exhibit heterogeneous annotation quality, category imbalance, and limited acoustic diversity. Curated public corpora provide relatively structured supervision but often underrepresent spontaneous NVVs and realistic recording conditions, whereas in-the-wild media recordings contain richer vocal behaviors but lack reliable annotations. These issues are particularly important for NVVSpeech Challenge Track~1\footnote{
  \url{https://nvvspeech-challenge.github.io}
}, which requires a bilingual tagged transcript jointly capturing lexical content, NVV identity, and transcript-relative placement~\cite{yang2026wesr,xue2026nvvsuperbench,ni2026nv}. Consequently, effective systems must address both NVV-aware sequence modeling and the construction of reliable, acoustically diverse supervision.

To this end, we develop an NVV-aware ASR system based on Whisper-medium together with a source-adaptive data curation strategy. NVV-Aware Whisper represents the 16 challenge NVV categories through checkpoint-compatible vocabulary remapping, enabling lexical tokens and inline NVV tags to be jointly generated within a single autoregressive sequence. For data construction, existing public corpora are refined through acoustic augmentation and multimodal LLM filtering, while naturally occurring NVVs are mined from in-the-wild media through preprocessing and automated annotation. The resulting complementary data sources are combined for bilingual model training.

Our contributions are threefold:
(i) an NVV-aware Whisper formulation that jointly transcribes lexical content and inline NVV events through vocabulary remapping without modifying the original model dimensions;
(ii) a source-adaptive data curation strategy that refines existing bilingual NVV corpora while mining spontaneous NVVs from in-the-wild media recordings;
and (iii) a systematic evaluation under the official NVVSpeech Track~1 protocol, where the proposed system improves FinalScore from 33.32 to 53.61, with controlled ablations quantifying the contributions of the major data-curation components.

\vspace*{-6pt}
\section{Method}

\subsection{Task Formulation}
Given an input waveform $\mathbf{x}$, the goal of NVV-aware ASR is to predict a tagged transcript $\mathbf{y}=(y_1,\ldots,y_T)$ that interleaves lexical tokens with inline nonverbal vocalization (NVV) tags. Track~1 considers a fixed 16-category NVV inventory: \nvvtag{breath}, \nvvtag{sniff}, \nvvtag{laugh}, \nvvtag{cry}, \nvvtag{cough}, \nvvtag{throat clearing}, \nvvtag{sneeze}, \nvvtag{sigh}, \nvvtag{gasp}, \nvvtag{snore}, \nvvtag{yawn}, \nvvtag{hum}, \nvvtag{moan}, \nvvtag{hiss}, \nvvtag{lipsmack}, and \nvvtag{burp}. The system therefore needs to jointly recognize lexical content, identify the NVV category, and place each NVV at the correct transcript-relative position. At inference time, only the audio is available; no visual stream, external event detector, or post-hoc NVV aligner is used.

\begin{figure*}[t]
  \centering
  \includegraphics[width=0.7\textwidth]{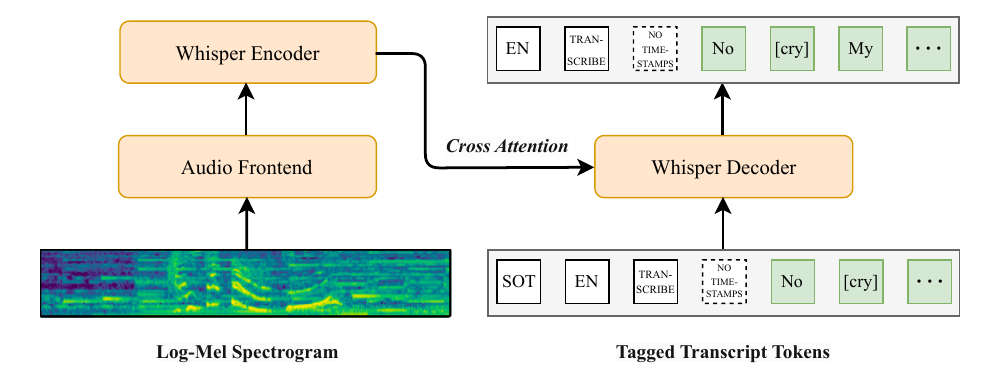}
  \vspace{-6pt}
  \caption{Overview of NVV-Aware Whisper. Whisper-medium is fully fine-tuned to jointly generate lexical tokens and inline NVV tags in a single autoregressive sequence, with timestamp prediction disabled. Flame icons denote trainable modules.}
  \label{fig:model}
\end{figure*}

\vspace*{-4pt}
\subsection{NVV-Aware Whisper}

Figure~\ref{fig:model} illustrates the proposed NVV-Aware Whisper framework.
We adopt Whisper-medium (769M parameters)~\cite{radford2023robust} as the backbone, an encoder--decoder Transformer~\cite{vaswani2017attention} whose autoregressive decoder supports inline tag generation unlike encoder-only ASR models~\cite{baevski2020wav2vec}.
The model is fully fine-tuned to generate lexical tokens and inline NVV tags within a single autoregressive sequence.
We retain Whisper's standard multilingual transcription prompt and disable timestamp prediction, since Track~1 requires transcript-relative NVV placement rather than absolute time alignment~\cite{DBLP:conf/interspeech/BainHHZ23}.

\textbf{NVV token representation.}
Since the original Whisper vocabulary does not contain dedicated tokens for the 16 NVV categories, we reuse 16 unused BPE entries and remap them to the canonical NVV tags.
The corresponding merge rules are removed so that each NVV label is represented as a single atomic token~\cite{papadourakis2021phonetically}.
This preserves the original vocabulary size and model dimensions, allowing the pretrained checkpoint to be directly reused without expansion.

\textbf{Joint lexical--NVV decoding.}
Lexical tokens and NVV tags share the same decoder output space and are generated autoregressively:
$p(\mathbf{y}\mid\mathbf{x})
=
\prod_{t=1}^{T}
p(y_t\mid y_{<t},\mathrm{Enc}(\mathbf{x}))$
Thus, lexical transcription, NVV recognition, and transcript-relative placement are handled within a unified sequence-generation framework without an additional event-classification or alignment module.
All model parameters are optimized using standard autoregressive cross-entropy.

\begin{table*}[t]
\centering
\caption{Composition of the training data at different curation stages.
Aug., LLM Filter, and Movie denote Data Augmentation, LLM-based Filtering, and the inclusion of in-the-wild media NVV data, respectively.
ZH/EN denotes the duration proportion of Mandarin and English, and NVV/h denotes the number of annotated NVV events per hour.}
\label{tab:data_stats}

\resizebox{\textwidth}{!}{%
\begin{tabular}{l l c c c r c r r r}
\toprule
\textbf{Source} &
\textbf{Dataset} &
\textbf{Aug.} &
\textbf{LLM Filter} &
\textbf{Movie} &
\textbf{Hours} &
\textbf{ZH / EN (\%)} &
\textbf{\#Utt.} &
\textbf{\#NVV} &
\textbf{NVV/h} \\
\midrule
\multirow{3}{*}{Public corpora}
& A: Raw Public NVV Data
& --
& --
& --
& 515.1
& 93.7 / 6.3
& 177{,}292
& 205{,}277
& 398.5 \\
& B: Augmented Public NVV Data
& \checkmark
& --
& --
& 763.0
& 93.6 / 6.4
& 262{,}400
& 303{,}800
& 398.3 \\
& C: Clean Public NVV Data
& \checkmark
& \checkmark
& --
& 545.7
& 91.6 / 8.4
& 192{,}014
& 222{,}150
& 407.1 \\
\midrule
In-the-wild media
& D: Movie NVV Data
& --
& --
& \checkmark
& 454.3
& 57.7 / 42.3
& 179{,}585
& 290{,}818
& 640.1 \\
\midrule
Public + media
& \textbf{E: Final NVV Data (C+D)}
& \checkmark
& \checkmark
& \checkmark
& \textbf{1{,}000.0}
& \textbf{76.2 / 23.8}
& \textbf{371{,}599}
& \textbf{512{,}968}
& \textbf{513.0} \\
\midrule
Speech-only corpora
& F: Tag-free ASR Data
& --
& --
& --
& 178.8
& 69.9 / 30.1
& 30{,}000
& \multicolumn{1}{c}{--}
& \multicolumn{1}{c}{--} \\
\midrule
All sources
& \textbf{Total Training Data (E+F)}
& \checkmark
& \checkmark
& \checkmark
& \textbf{1{,}178.8}
& \textbf{75.2 / 24.8}
& \textbf{401{,}599}
& \textbf{512{,}968}
& \multicolumn{1}{c}{--} \\
\bottomrule
\end{tabular}%
}
\end{table*}

\vspace*{-4pt}
\subsection{Source-Adaptive Data Curation}

Public NVV corpora and in-the-wild media recordings exhibit substantially different acoustic and annotation characteristics. Existing public corpora provide relatively structured bilingual supervision, but may suffer from limited acoustic diversity and inconsistent annotation quality. In contrast, movie and television recordings contain richer spontaneous NVVs, expressive delivery, speaker interactions, and realistic acoustic conditions, but lack reliable NVV annotations and require substantial preprocessing. We therefore adopt source-adaptive data curation strategies that refine existing public corpora while mining naturally occurring NVVs from in-the-wild media recordings.

\textbf{Public-corpus refinement.}
Raw Public NVV Data are assembled from the public bilingual corpora used by the released Track~1 baseline.
Since these datasets already provide NVV annotations, this branch focuses on improving their consistency, acoustic diversity, and label reliability through three steps:
\begin{itemize}[leftmargin=*, itemsep=2pt, parsep=0pt, topsep=0pt]

    \item \textit{Label normalization:}
    Annotations from different corpora are first mapped to the canonical 16-category NVV inventory.
    This unifies heterogeneous label conventions and ensures that all samples follow the same tagged-transcript format used for model training and evaluation.

    \item \textit{Acoustic augmentation:}
    Using noise and speed perturbation~\cite{ko2015audio,chen2019rare}, we add stratified waveform copies to improve acoustic diversity and alleviate tag imbalance, down-weighting dominant tags (e.g., \nvvtag{breath}, \nvvtag{laugh}) while upsampling tail tags.
Each selected utterance receives additive white Gaussian noise, speed perturbation, or both (40\%/40\%/20\%), with SNR uniform in $[20,35]$\,dB and speed factor in $[0.9,1.1]$; perturbed versions are appended as additional samples.
Mild ranges limit distortion of short or low-energy NVV cues such as \nvvtag{breath}, \nvvtag{sniff}, and \nvvtag{lipsmack}.

    \item \textit{Multimodal LLM filtering:}
    Augmentation does not correct unreliable labels; each waveform and tagged transcript are submitted to Gemini~2.5~Pro~\cite{comanici2025gemini} for accept-or-reject filtering.
    A sample is retained only when the annotated NVVs are judged to be acoustically present, correctly categorized, and consistent with the lexical transcript.
    Samples that fail these criteria are discarded rather than automatically repaired, thereby favoring annotation reliability over data quantity.
\end{itemize}
The retained samples constitute Clean Public NVV Data, providing relatively controlled bilingual supervision with improved acoustic diversity and label consistency.

\textbf{In-the-wild media data mining.}
Movie and television recordings provide spontaneous NVVs under substantially more diverse speakers, speaking styles, conversational contexts, and acoustic conditions than curated public corpora.
These recordings are therefore used to complement public data with naturally occurring NVV timing and speech--NVV co-occurrence patterns.
Unlike the public-data branch, no additional waveform-level augmentation is applied, so that subtle acoustic cues in the original recordings are preserved.
Raw Movie Data are converted into NVV-aware training samples through the following stages:
\begin{itemize}[leftmargin=*, itemsep=2pt, parsep=0pt, topsep=0pt]

    \item \textit{Audio preprocessing:}
    Audio tracks are first extracted using FFmpeg.
    MossFormer2-SE-48K~\cite{zhao2024mossformer2} is optionally applied to recordings with strong acoustic interference, while Silero-VAD~\cite{team2021silero} identifies speech-active regions and removes long non-speech intervals.
    This stage converts long-form media recordings into cleaner candidate regions for subsequent transcription and annotation.

    \item \textit{Transcription and segmentation:}
    Volcengine Doubao ASR~2.0~\cite{bai2024seed} is used to obtain preliminary sentence-level transcripts together with speaker information and timestamps.
    The ASR boundaries are combined with VAD regions using a 0.1\,s merge gap to reduce fragmented segmentation.
    Long recordings are then divided into utterances compatible with Whisper's 30\,s input window.
    The resulting transcript provides lexical scaffolding for subsequent NVV annotation rather than serving as the final training target.

    \item \textit{Automated NVV annotation:}
    Each segmented waveform and its preliminary transcript are jointly provided to Gemini~2.5~Pro, which produces a refined transcript containing both lexical content and inline NVV tags.
    In contrast to the accept-or-reject filtering used for public corpora, this stage performs generative annotation: NVV tags can be inserted, removed, corrected, or relocated according to the acoustic evidence.
    This allows previously unannotated media recordings to be converted into structured NVV-aware training examples while retaining naturally occurring event timing and contextual interactions.
\end{itemize}
The resulting samples constitute Movie NVV Data and complement the public corpora with spontaneous and acoustically diverse NVV realizations.

\textbf{Complementary data aggregation.}
Finally, Clean Public NVV Data and Movie NVV Data are combined to construct Final NVV Data.
The former contributes relatively controlled bilingual supervision, while the latter provides more spontaneous NVV realizations and diverse acoustic contexts.
We additionally mix tag-free ASR utterances into the training corpus as explicit speech-only negative examples.
These samples help preserve lexical transcription capability and discourage the decoder from generating NVV tags when no corresponding event is present.
Together, the three data components provide complementary supervision for joint lexical and NVV decoding.

\vspace*{-7pt}
\section{Experiments}

\vspace*{-7pt}
\subsection{Experimental Setup}
\vspace*{-3pt}
\textbf{Dataset}
Table~\ref{tab:data_stats} summarizes the composition of the final training corpus.
Final NVV Data combines Clean Public NVV Data and Movie NVV Data, which provide complementary supervision in terms of language coverage and naturally occurring NVV events.
After aggregation, it contains 762.0\,h of Mandarin and 238.0\,h of English speech, comprising 371,599 utterances and 512,968 annotated NVV events.
An additional 178.8\,h of tag-free ASR data (125.0\,h Mandarin / 53.8\,h English) is included as speech-only supervision, yielding a total training set of 1,178.8\,h and 401,599 utterances.
Evaluation is conducted on the official Final Stage test set of 1,946 utterances (985 Mandarin and 961 English) using the official Track~1 scorer.

\vspace*{-1pt}
\textbf{Systems}
All systems—including the baseline, proposed system, and ablation variants—adopt the same NVV-Aware Whisper architecture, training configuration, and decoding strategy.
They also share identical tag-free ASR data (178.8,h; Table~\ref{tab:data_stats}, Row~F) as fixed speech-only supervision, differing only in their NVV-labeled training corpus.
The \textbf{baseline} combines this fixed tag-free subset with raw public NVV data, following the official Track~1 baseline setting.
The \textbf{proposed} system preserves the same model, optimization recipe and tag-free ASR data, while replacing raw public NVV data with final NVV data (clean public NVV data \(+\) movie NVV data).
Ablation models modify individual NVV data-curation components with all other settings fixed, isolating the contributions of our NVV corpus construction pipeline.

\textbf{Parameters}
All systems are obtained by full-parameter fine-tuning of Whisper-medium using AdamW~\cite{loshchilov2017decoupled} with a cosine learning-rate schedule. The learning rate is set to $1\times10^{-5}$ with 500 warm-up steps, and the effective batch size is 352. Training uses bfloat16 precision and DeepSpeed ZeRO-2~\cite{rajbhandari2020zero} on eight NVIDIA RTX~4090 GPUs. Absolute timestamp prediction is disabled during both training and inference. At inference time, the Whisper language prompt is set according to the official language label of each utterance, and no sample-specific post-processing or external NVV detector is applied.

 \textbf{Evaluation Metrics}
Following the official Track~1 protocol, we evaluate NVV recognition, transcript-relative placement, and tagged transcription jointly. For each language, the Track1Score combines event-level micro-F1 ($\mathrm{F1_{micro}}$), multi-event normalized tag distance (mNTD), and tagged-transcript error ($\mathrm{Err_{tagged}}$):
\begin{equation}
\begin{split}
Score = 100\,[&
0.70\,\mathrm{F1}_{\mathrm{micro}}
+ 0.20(1-\mathrm{mNTD}) \\
&+ 0.10\left(1-\min(\mathrm{Err}_{\mathrm{tagged}},1)\right)
].
\end{split}
\label{eq:track1}
\end{equation}


Here, $\mathrm{F1_{micro}}$ measures event recognition, mNTD evaluates the transcript-relative placement of NVV tags, and $\mathrm{Err_{tagged}}$ denotes tag-aware CER for Mandarin and tag-aware WER for English. The overall bilingual score is the average of the Mandarin and English Track1Scores.
\begin{equation}
FinalScore = (Score_{\mathrm{ZH}} + Score_{\mathrm{EN}}) / 2.
\label{eq:bilingual}
\end{equation}
Since $\mathrm{F1_{micro}}$ accounts for 70\% of the Track1Score, accurate NVV recognition is the dominant factor, while tag placement and tagged-transcript accuracy provide complementary measures of localization and lexical fidelity.

\begin{table}[!t]
  \caption{Main results.
  Text-only is the corresponding CWE or WER after removing all NVV tags;
  }
  \label{tab:main}
  \centering
  \resizebox{\columnwidth}{!}{%
  \begin{tabular}{llcccccc}
    \toprule
    \textbf{System} & \textbf{Lang.} &
    \textbf{F1\_micro} $\uparrow$ & \textbf{mNTD} $\downarrow$ &
    \textbf{Err\_tagged} $\downarrow$ & \textbf{Text-only} $\downarrow$ &
    \textbf{Score} $\uparrow$ & \textbf{FinalScore} $\uparrow$\\
    \midrule
    \multirow{2}{*}{Baseline} & ZH & 0.3325 & 0.6602 & 60.24 & 48.15 & 34.05 & \multirow{2}{*}{33.32} \\
                              & EN & 0.2964 & 0.7459 & 32.38 & 25.58 & 32.59 & \\
    \midrule
    \multirow{2}{*}{Proposed} & ZH & 0.5666 & 0.4443 & 53.22 & 47.38 & 55.45 & \multirow{2}{*}{\textbf{53.61}} \\
                              & EN & 0.5081 & 0.4357 & 50.91 & 34.98 & 51.76 & \\
    \bottomrule
  \end{tabular}}
\end{table}

\vspace*{-5pt}
\subsection{Main Results and Analysis}
\vspace*{-5pt}
Table~\ref{tab:main} compares Baseline and Proposed systems.
Replacing Raw Public NVV Data with the final training corpus increases bilingual FinalScore from 33.32 to 53.61, a 20.29-point absolute gain.
This improvement is consistent across languages: Mandarin Track1Score increases from 34.05 to 55.45 and English from 32.59 to 51.76.
The largest gains occur in NVV recognition and placement: Mandarin $\mathrm{F1_{micro}}$ increases from 0.3325 to 0.5666 while mNTD decreases from 0.6602 to 0.4443; English $\mathrm{F1_{micro}}$ increases from 0.2964 to 0.5081 while mNTD decreases from 0.7459 to 0.4357.
Thus, the source-adaptive data curation strategy substantially improves NVV event recognition and transcript-relative localization.

To disentangle lexical recognition from NVV-related errors, Table~\ref{tab:main} additionally reports text-only CER/WER without NVV tags.
For Mandarin, text-only CER remains nearly unchanged (48.15\%$\rightarrow$47.38\%), whereas tagged CER improves more substantially from 60.24\% to 53.22\%.
This suggests the Mandarin gain stems primarily from improved NVV recognition and placement with lexical ASR largely preserved.
For English, however, text-only WER increases from 25.58\% to 34.98\%, indicating genuine lexical degradation besides NVV-related errors.
Tagged WER increases even more sharply, from 32.38\% to 50.91\%, showing that NVV insertion, substitution, and placement errors further contribute to tagged-transcript error.

Despite this English transcription degradation, substantial improvements in $\mathrm{F1_{micro}}$ and mNTD dominate the overall score because these NVV-oriented metrics account for 90\% of Track1Score.
The two languages therefore exhibit different behaviors: Mandarin achieves substantially better NVV recognition and localization while largely preserving lexical accuracy, whereas English obtains comparable NVV gains at the cost of reduced transcription fidelity.
Notably, although Movie NVV Data substantially increase English coverage, the final training corpus remains Mandarin-dominant (887.0 vs.\ 291.8\,h; Table~\ref{tab:data_stats}), suggesting that improving language balance and English lexical robustness remains important for further improvement.

\begin{table}[!tb]
\caption{Per-class F1 of Baseline ($F1_{\mathrm{B}}$) and Proposed ($F1_{\mathrm{P}}$).
}

\label{tab:per_class_f1}
\centering
\footnotesize
\setlength{\tabcolsep}{4pt}
\renewcommand{\arraystretch}{1.05}

\resizebox{\columnwidth}{!}{%
\begin{tabular}{@{}l cc cc p{2.5em} l cc cc@{}}
\toprule
\multirow{2}{*}{\textbf{Tag}} & \multicolumn{2}{c}{\textbf{EN}} & \multicolumn{2}{c}{\textbf{ZH}}
& & \multirow{2}{*}{\textbf{Tag}} & \multicolumn{2}{c}{\textbf{EN}} & \multicolumn{2}{c}{\textbf{ZH}} \\
\cmidrule(lr){2-3} \cmidrule(lr){4-5} \cmidrule(lr){8-9} \cmidrule(lr){10-11}
& \textbf{F1$_{\mathrm{B}}$ $\uparrow$} & \textbf{F1$_{\mathrm{P}}$ $\uparrow$}
& \textbf{F1$_{\mathrm{B}}$ $\uparrow$} & \textbf{F1$_{\mathrm{P}}$ $\uparrow$}
& &
& \textbf{F1$_{\mathrm{B}}$ $\uparrow$} & \textbf{F1$_{\mathrm{P}}$ $\uparrow$}
& \textbf{F1$_{\mathrm{B}}$ $\uparrow$} & \textbf{F1$_{\mathrm{P}}$ $\uparrow$} \\
\midrule
breath          & 0.051 & \textbf{0.330} & 0.073 & \textbf{0.237} & &
gasp            & 0.171 & \textbf{0.224} & 0.047 & \textbf{0.288} \\
sniff           & 0.145 & \textbf{0.532} & 0.168 & \textbf{0.679} & &
snore           & 0.286 & \textbf{0.457} & \textbf{0.904} & 0.856 \\
laugh           & 0.158 & \textbf{0.525} & 0.500 & \textbf{0.615} & &
yawn            & 0.338 & \textbf{0.703} & 0.616 & \textbf{0.809} \\
cry             & 0.185 & \textbf{0.481} & 0.249 & \textbf{0.454} & &
hum             & 0.311 & \textbf{0.713} & 0.718 & \textbf{0.751} \\
cough           & 0.314 & \textbf{0.620} & 0.352 & \textbf{0.608} & &
moan            & 0.165 & \textbf{0.340} & 0.082 & \textbf{0.473} \\
throat clearing & 0.483 & \textbf{0.803} & 0.269 & \textbf{0.739} & &
hiss            & \textbf{0.660} & 0.625 & 0.098 & \textbf{0.576} \\
sneeze          & 0.338 & \textbf{0.676} & 0.209 & \textbf{0.790} & &
lipsmack        & \textbf{0.540} & 0.423 & 0.231 & \textbf{0.458} \\
sigh            & 0.019 & \textbf{0.498} & 0.395 & \textbf{0.597} & &
burp            & 0.161 & \textbf{0.742} & 0.930 & \textbf{0.971} \\
\bottomrule
\end{tabular}%
}
\end{table}

\vspace*{-7pt}
\subsection{Per-Class Analysis}

Table~\ref{tab:per_class_f1} lists event-level F1 for all 16 NVV categories.
The Proposed system improves 14 of 16 categories in English and 15 of 16 in Mandarin, indicating that the overall gain is broadly distributed rather than dominated by a small subset of NVVs.
Particularly large improvements are observed for English \nvvtag{burp} ($0.161\rightarrow0.742$), \nvvtag{sigh} ($0.019\rightarrow0.498$), and \nvvtag{hum} ($0.311\rightarrow0.713$), as well as Mandarin \nvvtag{sneeze} ($0.209\rightarrow0.790$), \nvvtag{sniff} ($0.168\rightarrow0.679$), \nvvtag{hiss} ($0.098\rightarrow0.576$), and \nvvtag{throat clearing} ($0.269\rightarrow0.739$).
These consistent gains across diverse categories suggest that the proposed data curation improves NVV coverage beyond a few frequent event types.

Several categories nevertheless remain challenging.
English \nvvtag{gasp}, \nvvtag{breath}, and \nvvtag{moan} achieve F1 scores of only 0.224, 0.330, and 0.340, respectively, despite Baseline improvements.
Mandarin shows similar difficulty for \nvvtag{breath} and \nvvtag{gasp}, at 0.237 and 0.288.
This indicates some fine-grained NVV categories remain substantially harder to recognize and may require targeted supervision.

Performance is not uniformly improved across all classes.
English \nvvtag{hiss} and \nvvtag{lipsmack} decrease from 0.660 to 0.625 and from 0.540 to 0.423, respectively, while Mandarin \nvvtag{snore} decreases slightly from 0.904 to 0.856.
Moreover, several categories exhibit strong language-dependent behavior; for example, Mandarin \nvvtag{burp} is already near saturation in the Baseline (0.930), whereas English \nvvtag{burp} improves substantially from 0.161 to 0.742.
Overall, the per-class results demonstrate broad improvements from the proposed training data while revealing residual category- and language-specific weaknesses that are obscured by the aggregate micro-F1 score.

\begin{table}[t]
  \caption{Ablation of different NVV training-data configurations. All systems additionally share the same fixed tag-free ASR data.}
  \label{tab:ablation}
  \centering
  \footnotesize
  \renewcommand{\arraystretch}{1.05}

  \resizebox{\columnwidth}{!}{
  \begin{tabular}{l c c c}
    \toprule
    \textbf{Training Data} &
    \textbf{ZH Score} $\uparrow$ &
    \textbf{EN Score} $\uparrow$ &
    \textbf{FinalScore} $\uparrow$ \\
    \midrule

    A: Raw Public NVV Data & 34.05 & 32.59 & 33.32 \\
    B: Augmented Public NVV Data & 38.96 & 36.16 & 37.56 \\
    C: Clean Public NVV Data & 41.08 & 37.46 & 39.27 \\
    D: Movie NVV Data & 54.21 & 38.47 & 46.34 \\
    E: Final NVV Data & \textbf{55.45} & \textbf{51.76} & \textbf{53.61} \\
    \bottomrule
  \end{tabular}
  }
\end{table}

\vspace*{-5pt}
\subsection{Data-Curation Ablation}
\label{sec:ablation}

Table~\ref{tab:ablation} evaluates different NVV training-data configurations while keeping the model, optimization settings, and tag-free ASR data fixed.
Along the public-corpus refinement path, Data Augmentation improves the final score from 33.32 to 37.56 (+4.24), while subsequent LLM-based Filtering further raises it to 39.27 (+1.71).
The consistent gains in both Mandarin and English confirm that augmentation and filtering provide complementary improvements to the public NVV data.

Movie NVV Data alone achieves a substantially higher final score of 46.34 than Clean Public NVV Data (39.27), although its performance is markedly stronger in Mandarin than in English (54.21 vs.\ 38.47).
Combining the two sources yields the best overall result of 53.61, corresponding to gains of +14.34 over Clean Public NVV Data and +7.27 over Movie NVV Data alone.
Notably, the combination improves English from 38.47 to 51.76 while further increasing Mandarin from 54.21 to 55.45.

Overall, the ablation demonstrates that public-corpus refinement and in-the-wild media data are complementary: the former provides incremental gains through augmentation and filtering, while the latter offers stronger standalone performance; their combination achieves the best and most balanced bilingual result.

\vspace*{-5pt}
\section{Conclusion}

We presented a bilingual NVV-aware ASR system that combines joint lexical–NVV decoding with source-adaptive data curation. Under the official NVVSpeech Track~1 protocol, the proposed system improves FinalScore from 33.32 to 53.61, with ablations confirming individual benefits of Data Augmentation, LLM-based Filtering, and Movie NVV Data. Notably, per-class and text-only analyses show broad NVV gains but reveal persistent challenges in fine-grained categories and English transcription robustness, indicating these issues remain unresolved. Future work will focus on better language balance and weaker trade-off between NVV recognition and lexical accuracy.

\bibliographystyle{IEEEtran}
\bibliography{cyabib}

\end{document}